\documentclass[a4paper,11pt]{article}
\usepackage{pos}
\usepackage{xcolor}
\usepackage{natbib}

\renewenvironment{thebibliography}[1]{
  \begin{oldthebibliography}{#1}
    \setlength{\itemsep}{0.35em}
    \setlength{\parskip}{0em}
}
{
  \end{oldthebibliography}
}

\usepackage{booktabs, tabularx}

\setcitestyle{author,open={[},close={]}}

\title{Deep Learning Transient Detection with VERITAS}
 \ShortTitle{Deep Learning Transient Detection with VERITAS}

\author*[a]{Konstantin Pfrang}

\affiliation[a]{Deutsches Elektronen Synchrotron DESY,\\
  Platanenallee 6, 15738 Zeuthen, Germany}

\forColl{VERITAS} 

\emailAdd{konstantin.pfrang@desy.de}

\abstract{Ground-based $\gamma$-ray observatories, such as the VERITAS array of imaging atmospheric Cherenkov telescopes, provide insight into very-high-energy (VHE, $\mathrm{E}>100\,\mathrm{GeV}$) astrophysical transient events. Examples include the evaporation of primordial black holes, gamma-ray bursts and flaring blazars. Identifying such events with a serendipitous location and time of occurrence is difficult. Thus, employing a robust search method becomes crucial. An implementation of a transient detection method based on deep-learning techniques for VERITAS will be presented. This data-driven approach significantly reduces the dependency on the characterization of the instrument response and the modelling of the expected transient signal. The response of the instrument is affected by various factors, such as the elevation of the source and the night sky background. The study of these effects allows enhancing the deep learning method with additional parameters to infer their influences on the data. This improves the performance and stability for a wide range of observational conditions.
We illustrate our method for an historic flare of the blazar BL Lac that was detected by VERITAS in October 2016. We find a promising performance for the detection of such a flare in timescales of minutes that compares well with the VERITAS standard analysis.}

\FullConference{37$^{\rm{th}}$ International Cosmic Ray Conference (ICRC 2021)\\
		July 12th -- 23rd, 2021\\
		Online -- Berlin, Germany}

\begin{document}
\maketitle

\section{Introduction}
The detection of astrophysical transient events at very-high energies provides insight into different fundamental phenomena. 
These include previously detected transient phenomena such as gamma-ray bursts \cite{grbHESS2019} or flaring blazars \cite{bllacflare2018} and hypothesized transient phenomena such as the evaporation of primordial black holes \cite{carrPBH2010}.
The Very Energetic Radiation Imaging Telescope Array System (VERITAS) is an array of four imaging atmospheric Cherenkov telescopes (IACTs) located at the Fred Lawrence Whipple Observatory in southern Arizona (31 40\,N, 110 57\,W,  1.3\,km a.s.l.) and is sensitive to sources of very-high-energy (VHE) photons above $100$\,GeV with a strength of $1\%$\,Crab Units (C.U.) in $\sim25$\,h \cite{holder2011}.
VERITAS can detect VHE astrophysical transient sources within its field of view (FoV) of diameter of $3.5^\circ$ \cite{holder2006}.

IACTs detect an inevitable rate of background events, which pass the nominal selection cuts.
The background includes $\gamma$-like cosmic-ray showers, electrons, positrons, and, to a lesser extent, diffuse $\gamma$ rays.
An astrophysical source may cause an enhanced event rate that over this background rate.
Transient sources, cause a temporary increase of the observed event rates.
These can appear at any location in the FoV and at any time. 
The observed rates change depending on the observation conditions. 
Thus, the changes in the instrument response need to be taken into account for a robust method to identify transients in the data.
In this work, the approach proposed for IACTs in \cite{sadeh2020} is implemented for VERITAS.
This deep-learning-based transient detection technique extracts information about the changes of the event rates from the data themselves.
Being fully data-driven, it is insensitive to uncertainties on the modelling of the instrument and can operate across the entire FoV.
It represents the first application of this method to real IACT data.

\section{Deep Learning Transient Detection}
Machine learning techniques are widely used in astronomy \cite{MLastro2010}. Among them, deep learning (DL) has proven its power in various astronomical applications (e.g., see \cite{DL2019review}). 
It is based on deep and wide artificial neural networks (ANNs) that are able to represent complex models \cite{DL2015general}. 
Recurrent Neural Networks (RNNs) are a type of ANNs that include cyclic connections that are especially suited to work with sequential data, such as time series.
We use the framework presented in \cite{sadeh2020} that can be decomposed into a pair of encoder and decoder stages.
As input, it takes a total of $\tau_{\mathrm{RNN}}=\tau_{\mathrm{enc}}+\tau_{\mathrm{dec}}$ time steps.
At each step, event counts $N_{\mathrm{evt}}$ in $\eta$ energy bins are given as inputs to the network.
The $\tau_\mathrm{enc}$ steps of the encoder represent the background to a possible transient signal.
Subsequently, the signal is searched for in the ensuing $\tau_{\mathrm{dec}}$ decoder steps.
Each time step of the RNN consists of a long short-term memory (LSTM) unit \cite{LSTM1997}, comprised of a layer of 64 hidden units.
The RNN is trained with sequences of time series which do not contain any signals (either in the encoder or in the decoder intervals). 
It thus learns to predict the expected background counts $B(\tau_{\mathrm{dec}},\eta)$.
If a transient signal is present as part of the decoder interval, the total observed counts $S(\tau_{\mathrm{dec}},\eta)$ are potentially increased.
In this work, we use a different test statistic (TS) than the one used for \cite{sadeh2020}, who assumed that the inputs to the network followed Poissonian statistics.
We instead define our TS as
\begin{align}
    \mathrm{TS(\eta)} = \frac{S(\tau_{\mathrm{dec}}, \eta) - B(\tau_{\mathrm{dec}}, \eta)}{\sqrt{\left|B(\tau_{\mathrm{dec}}, \eta)\right| +1}}.
\end{align}
The one in the denominator is added to avoid numerical division by zero.
The TS is independently computed for the event counts in each energy bin. 
As the training data consists exclusively of background sequences, it can be used to map TS values into a $p$-value for detection. 
We define the combined TS as the average of $|\log(p)|$ over all features and map it to a combined $p$-value itself. 

\section{Data Selection and Preparation}
This work utilizes VERITAS data recorded from 2012 to 2020. 
To learn the changes in the background rates, a dataset that does not include uncontrolled systematic effects is required.
Possible systematics include hardware issues or the effects of bad weather, e.g., by clouds in the field of view.
As these result in a drop of the event rates, they potentially degrade the reliability of a search for transients.

For the generation of the training dataset,
we apply a selection on data-taking runs, based on existing data quality monitoring results and basic filters.
We consider data that was taken while operating the full four-telescope array at zenith angles below $45^\circ$ with nominal high voltage settings and without recorded hardware issues.
Manual data quality checks for a large dataset are time-consuming and a possible source of inconsistency.
To avoid this, we deploy an automatic pipeline, that assesses the data quality after the initial selection.
We investigate time series data, such as the array trigger (L3) rate and temperature of the sky for systematic effects. 
Among these are spikes or drops in the trigger rate e.g., due to a short flash of light in the camera, clouds in the FoV, and changes correlated with the amount of night sky background (NSB).
When such substantial systematics are detected, time cuts are applied and the corresponding interval in the data is discarded.
We validate the remaining intervals for consistency with a constant trigger rate to indicate a stable performance of VERITAS, where a final time cut may also be applied.
Following this procedure, we obtain a training dataset with an integrated $\sim{2,730}\,\mathrm{hours}$ of observations.

The data are processed using one of the standard VERITAS analysis packages \cite{EDmaier}.
We start from a list of events following the standard event reconstruction, e.g., having reconstructed energies and directions of arrival.
\begin{figure}[htbp]
    \centering
    \includegraphics[width=0.7\textwidth]{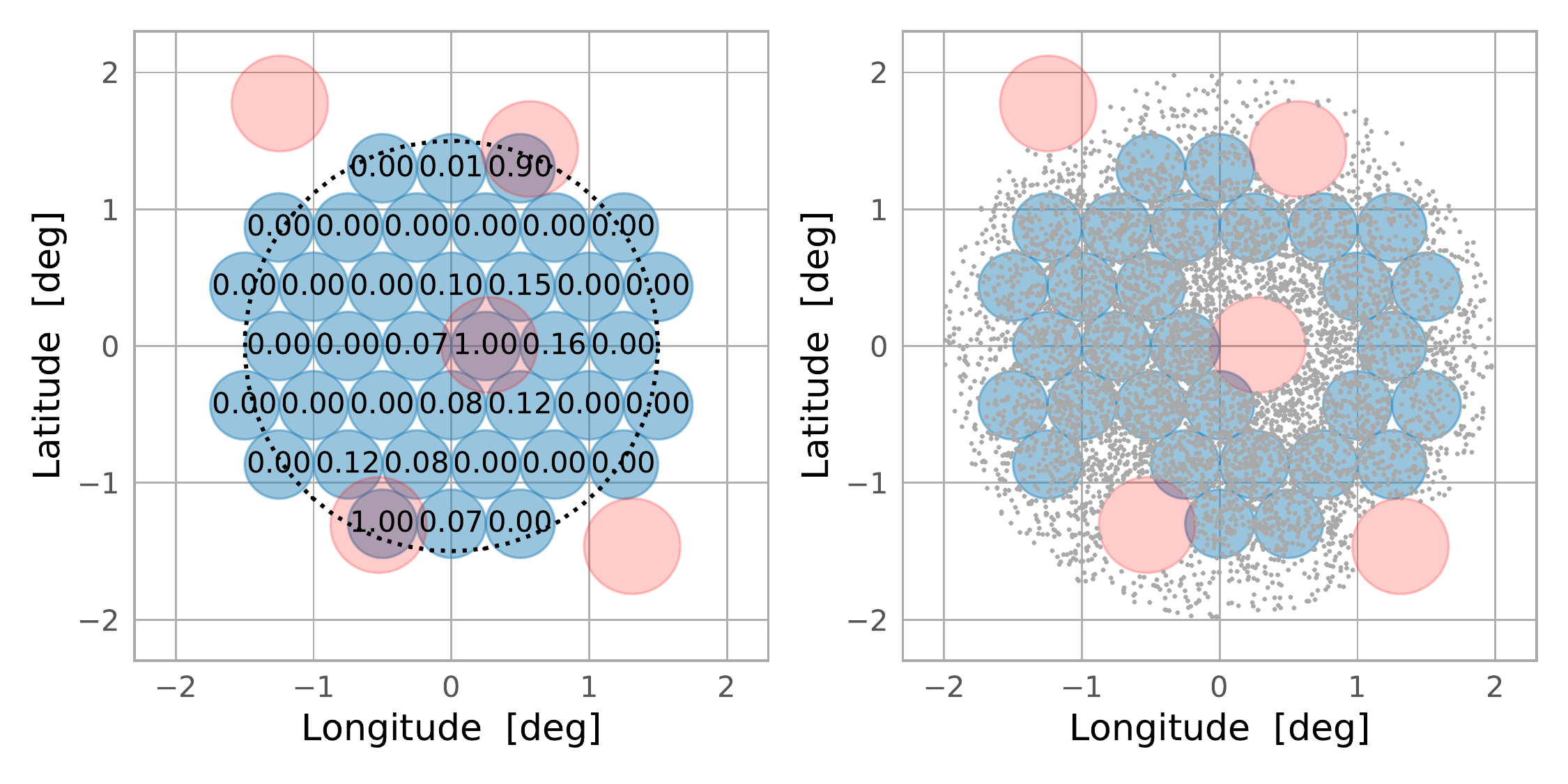}
    \caption{Example for the definition of ROIs in the VERITAS FoV. \textit{Left:} The initial ROIs, shown in blue, are filling the inner $1.5^\circ$ relative to the center of the camera. The red circles indicate the defined exclusion regions due to bright stars or known $\gamma$-ray sources. The numbers in each ROI, are the fraction of area excluded. \textit{Right:} ROIs that  overlap by more than $10\%$ with an exclusion region are discarded. The small grey dots indicate the origin of the $\gamma$-like events. Events that fall into a surviving ROI are kept for the analysis.} 
    \label{fig:rois}
\end{figure}
For each run, we define our ROIs as non-overlapping circles, up to a distance of $1.5^\circ$ from the center of the FOV. 
See \autoref{fig:rois}.
We choose a radius for the ROIs of $0.25^\circ$. This conservatively covers the point spread function of VERITAS (uncertainty on event localisation), over our chosen event parameter range.
Bright stars and known VHE $\gamma$-ray sources in the FoV are masked with exclusion regions.
Events coming from these directions are not considered for the analysis.
The observations are binned into time steps of $30\,\mathrm{s}$ intervals. 
Time cuts that are determined by the automatic quality assessment are applied.
If more than $10\%$ of a ROI or a time step are masked, they are removed completely.
Otherwise we correct the measured events for the expected missing fraction, assuming uniform distributions within each ROI.
In each of the ROIs and time steps, the count of events is calculated in three energy bins, $100\,\mathrm{GeV}$ to $330\,\mathrm{GeV}$, $330\,\mathrm{GeV}$ to $1\,\mathrm{TeV}$, and $1\,\mathrm{TeV}$ to $100\,\mathrm{TeV}$. 

For $\gamma$-hadron separation (classification of events as non-$\gamma$-ray background), boosted decision trees (BDTs) are utilized \cite{bdt2017}.
These assign a score between $-1$ and $1$ to each event, which indicates how "$\gamma$-like" it appears. 
Depending on the zenith and energy bin, events with $\mathrm{BDT}\geq0.34$ to $\mathrm{BDT}\geq0.71$ are selected in the standard VERITAS analysis.
In this work, we deploy two different counting schemes to obtain the number of events $N_{\mathrm{evt}}$ per bin.
In the first approach, we do not apply the cut on the BDT-score, which was optimized for the traditional analysis.
Instead, only a loose $\mathrm{BDT-score}\,\geq-0.5$ cut is used to remove the bulk of events with the highest probability of being background.
We calculate a weighted sum of the remaining events.
The event weights are the BDT-scores remapped linearly between 0, corresponding to $\mathrm{BDT}=-0.5$, and 1, corresponding to $\mathrm{BDT}=1$. 
In the following we refer to it as \textit{weighted} counting scheme.
For the second approach, we only count the number of events that pass the traditional $\gamma$-hadron cuts (without applying weights).
Both approaches are utilized together, resulting in two count-metrics for each energy bin.
We therefore have a total of six inputs for each time step $\tau_{\mathrm{RNN}}$, which are mapped to six corresponding test statistics (that are combined into a final TS, as described above).

The training and calibration of the TS requires that the training data does not include transient signals.
Thus, for generation of the training dataset, we shuffle the timestamps and the polar angles in camera coordinates (assuming radial symmetry of the camera acceptance) of each event.
Exclusion regions and time cuts are preserved as part of the shuffling process.

\section{Auxiliary Parameters and Meta Bins}
VERITAS operates under a wide range of observation conditions, which affects the performance of the instrument.
Our DL data-driven approach learns to predict the different background rates from the training dataset.
Auxiliary parameters (in addition to event counts) are used to inform the RNN about potential systematics. 
We investigated parameters such as the \textit{L3 rate}, the \textit{NSB level}, the \textit{azimuth}, the secant of the \textit{zenith angle} $\sec(\theta)$, the reference observation time \textit{ref\_time}, the \textit{offset} of the ROI compared to the center of the camera, and the average number of images per event (\textit{multiplicity}).
An overview of the auxiliary parameters is provided in \autoref{tab:aux}.
The conclusions from our study are the following: 
$-$~The azimuth and $\sec(\theta)$ effectively account for the difference in rates due to the pointing of the instrument.
$-$~The reference time is crucial to characterize performance changes due to upgrades and aging effects (see also \cite{flux2021icrc}).
$-$~The offset distance accounts for the difference in the radial acceptance.
$-$~Finally, the multiplicity in the different energy bins contributes to identifying differences in the energy threshold that affects the rate of low-energy events.
\begin{table}[htbp]
    \caption{Auxiliary parameters and meta bins for prediction of the background counts.}
    \label{tab:aux}
    \centering
    \begin{tabularx}{0.99\textwidth}{m{2.1cm}m{6.cm}m{0.6cm}m{4.4cm}}
    \toprule[1.5pt]
    \textit{Parameter} & \textit{Description} & \textit{Used} & \textit{Meta bins}\\
    \midrule
        l3\_mean & Mean L3 trigger rate during time step & No &{--}\\
        nsb\_level & Mean charge in camera indicating NSB & No & {--}\\
        azimuth & Azimuth of pointing position & Yes & North, South \\
        $\sec(\theta)$ & secant of the pointing zenith angle & Yes & [1.0, 1,08), [1.08, 1.16), [1.16, 1.24), [1.24, 1.4)\\
        ref\_time & Time after August 1, 2012 in years & Yes & 2012-13, 2013-14, 2014-20\\
        offset & Distance of ROI to camera center & Yes & 6 bins, maximum $10\%$ difference in radial acceptance\\
        multiplicity($\eta$) & Average number of images at each time step and energy bin & Yes & {--}\\
    \bottomrule[1.5pt]
 
    \end{tabularx}
\end{table}

As the expected background counts change, the interpretation of the TS can be adjusted.
We introduce meta bins to split the observations into chunks in which the observed background rates are mostly stable.
The study to determine the auxiliary parameters revealed that the most critical parameters are the azimuth, $\sec(\theta)$, ref\_time, and the offset.
In total, we define 144 bins that are summarized in \autoref{tab:aux}.


\section{Detection of BL Lac Flare}
In its low state, the blazar BL Lac is not detectable in VERITAS over the timescales investigated in this work (few minutes).
However, in the past it proved to be a strong transient source, exhibiting flaring periods.
We investigate the BL Lac flare which occurred during October 2016 that reached a flux of $\sim1.8\,\mathrm{C.U.}$ above $200\,\mathrm{GeV}$ in the 4-minute-binned VERITAS light curve \cite{bllacflare2018}.
In this work, we consider the 30 minute run that corresponds to the highest flux in the 30-minute-binned light curve (see Figure 1 of \cite{bllacflare2018}). We investigate the potential to detect such a flare with the DL transient detection method.

\begin{figure}[htbp]
    \centering
    \includegraphics[width=\textwidth]{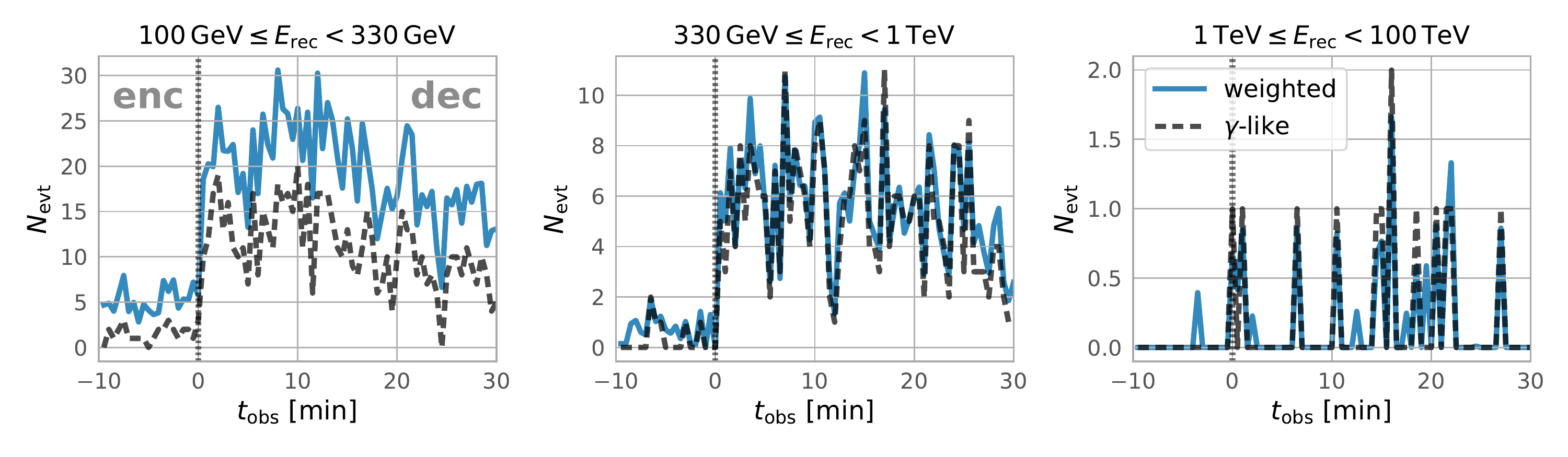}
    \caption{Number of events $N_{\mathrm{evt}}$ as function of the time $t_\mathrm{obs}$ in the generated light curve of BL Lac in three energy bins. The transition between the encoder and the decoder interval is at $t_\mathrm{obs}=0$.
    The solid blue curves represent the counts obtained from the weighted sum after the loose BDT cut. The dashed black lines are the counts of events after the $\gamma$-hadron separation. The encoder steps are sampled from a shuffled low state dataset and the decoder shows the counts during the flare.}

    \label{fig:lc}
\end{figure}

\begin{figure}[htbp]
    \centering
    \includegraphics[width=1.\textwidth]{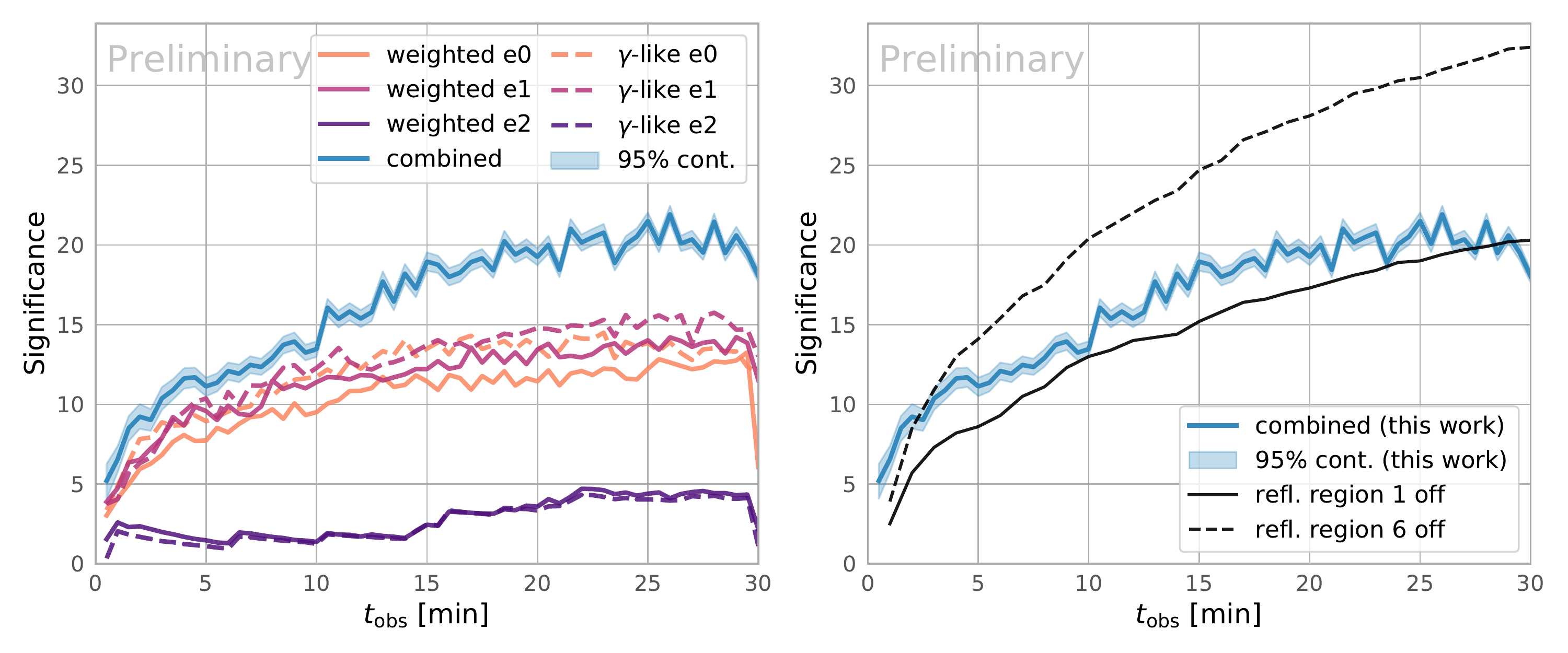}
    \caption{Significance of detection as function of the observation time of the flare $t_{\mathrm{obs}}$. 
    The calculation is repeated 1000 times while resampling the encoder steps.
    We show the medium and the $95\%$ containment interval in blue. 
    The left graph shows the individual contribution to the significance of all 6 parameters.
    Solid lines represent the weighted counts after the loose BDT cuts and the dashed are counts of $\gamma$-like events. The individual colors show the three energy bins.
    The black reference curves in the right are derived from the VERITAS standard analysis above $100\,\mathrm{GeV}$ using the reflected region background with 1 and 6 off regions respectively.}
    \label{fig:sig_time}
\end{figure}

The decoder steps, $\tau_{\mathrm{dec}}$, in which the signal is searched are filled with data obtained during the flaring period.
The DL transient method requires inputs for the encoder steps, $\tau_{\mathrm{enc}}$, that represent the background event rates.
As the source is in the flaring state, we generate the encoder counts by sampling from shuffled BL Lac observations during a low state, taken under similar observing conditions.
Generally, the padding for the encoder steps can also be extracted directly from the training dataset, within the same meta bins.
The light curves for both counting schemes are shown in \autoref{fig:lc}.
\\
We iteratively increase the number of decoder steps from 1 to 60, investigating the change in detection significance as a function of the observation time $t_\mathrm{obs}$ of the flare.
The results are shown as the blue curves in \autoref{fig:sig_time}.
In this study, the padding of the encoder samples is repeated 1000 times in order to derive the corresponding uncertainties ($95\%$ containment).
On the left graph, we investigate the individual contributions of the 6 input features to the combined significance. 
One may observe that the significance for the two different counting metrics are comparable. 
Overall the $\gamma$-like counts perform slightly better on longer timescales.
This reflects the higher signal to noise ratio, as a larger fraction of the background is discarded.
For short timescales and at highest energies, the weighted counting metric is slightly more informative for the network, given that this regime is dominated by statistical uncertainties.

The TS is calculated and mapped to p-values for each of the six input counts independently.
This allows our method to also perform well in cases where only a a subset of the features becomes particularly important.
For example a very hard source spectrum could lead to a large enhancement of the $\gamma$-ray rate in the highest energy bin.
As a very low background rate is expected in this case, the related significance would be comparably higher.
On the right side of \autoref{fig:sig_time}, we compare the combined significance against the results obtained for $\mathrm{E}>100\,\mathrm{GeV}$ with the VERITAS standard analysis that uses equation 17 of \cite{lima}.
We use the \textit{reflected region} background method \cite{berge_bg}.
For this, \textit{off regions} with the same size as the target region are defined on a circle around the camera center. 
The background is contemporaneously determined from the number of $\gamma$-like events arriving from these regions.
The background estimation for the RNN is based on the encoder steps of one ROI, in an area corresponding to a single off region.
The significance obtained with our method is compatible with the results of the standard analysis throughout most of the run.
For short timescales, the reflected region method is limited by low background statistics.
On the other hand, the DL method predicts the counts based on previous observations, and so is more robust. 
Thus, our approach has the potential to outperform the standard method on short time scales.
This can be critical when dealing with transient signals that might have rapid variability or short timescales.

We also show for illustration results for the standard VERITAS analysis using six off regions.
A higher number of off regions lead to better estimation of the background, and correspondingly to higher detection significance.
In general, the definition of off regions in the FoV can be challenging for geometrical reasons or the presence of $\gamma$-ray sources.
Thus, this method is often used for detailed offline analyses instead for blind searches across the entire FoV.
In future studies, we will investigate alternative training schemes for our RNN, which might take advantage of larger background estimation areas. 
However, we also note that on short timescales, even given the addition of off regions, the DL approach performs slightly better in comparison.

\section{Summary}
In this study we present the first implementation of a deep learning based method to detect transient sources with VERITAS.
We developed a pipeline to perform an automatic data quality assessment and convert the event lists of the standard analysis to the inputs.
The required auxiliary parameters and meta bins were determined by studying their effects on the observed background rates.
As part of the anomaly detection approach presented in this work, the network learns to predict the expected background rates. 
A transient signal can be detected as a divergence from these predictions.
We present a preliminary choice of selected inputs to such a network.

We illustrate our methodology on an historical flare of the blazar BL Lac, which was observed during October 2016.
The generated time series correspond to a scenario of a possible follow-up observation.
Overall the results are compatible with the significance achieved by traditional detection methods, considering a single reflected background region.
The results for short timescales are promising, which can be critical for the fast detection of transient signals.

\section*{Acknowledgements}
This research is supported by grants from the U.S. Department of Energy Office of Science, the U.S. National Science Foundation and the Smithsonian Institution, by NSERC in Canada, and by the Helmholtz Association in Germany. This research used resources provided by the Open Science Grid, which is supported by the National Science Foundation and the U.S. Department of Energy's Office of Science, and resources of the National Energy Research Scientific Computing Center (NERSC), a U.S. Department of Energy Office of Science User Facility operated under Contract No. DE-AC02-05CH11231. We acknowledge the excellent work of the technical support staff at the Fred Lawrence Whipple Observatory and at the collaborating institutions in the construction and operation of the instrument. K. Pfrang acknowledges the support of the Young Investigators Program of the Helmholtz Association.

\bibliography{reference}

\clearpage \section*{Full Authors List: \Coll\ Collaboration}

\scriptsize
\noindent
C.~B.~Adams$^{1}$,
A.~Archer$^{2}$,
W.~Benbow$^{3}$,
A.~Brill$^{1}$,
J.~H.~Buckley$^{4}$,
M.~Capasso$^{5}$,
J.~L.~Christiansen$^{6}$,
A.~J.~Chromey$^{7}$, 
M.~Errando$^{4}$,
A.~Falcone$^{8}$,
K.~A.~Farrell$^{9}$,
Q.~Feng$^{5}$,
G.~M.~Foote$^{10}$,
L.~Fortson$^{11}$,
A.~Furniss$^{12}$,
A.~Gent$^{13}$,
G.~H.~Gillanders$^{14}$,
C.~Giuri$^{15}$,
O.~Gueta$^{15}$,
D.~Hanna$^{16}$,
O.~Hervet$^{17}$,
J.~Holder$^{10}$,
B.~Hona$^{18}$,
T.~B.~Humensky$^{1}$,
W.~Jin$^{19}$,
P.~Kaaret$^{20}$,
M.~Kertzman$^{2}$,
T.~K.~Kleiner$^{15}$,
S.~Kumar$^{16}$,
M.~J.~Lang$^{14}$,
M.~Lundy$^{16}$,
G.~Maier$^{15}$,
C.~E~McGrath$^{9}$,
P.~Moriarty$^{14}$,
R.~Mukherjee$^{5}$,
D.~Nieto$^{21}$,
M.~Nievas-Rosillo$^{15}$,
S.~O'Brien$^{16}$,
R.~A.~Ong$^{22}$,
A.~N.~Otte$^{13}$,
S.~R. Patel$^{15}$,
S.~Patel$^{20}$,
K.~Pfrang$^{15}$,
M.~Pohl$^{23,15}$,
R.~R.~Prado$^{15}$,
E.~Pueschel$^{15}$,
J.~Quinn$^{9}$,
K.~Ragan$^{16}$,
P.~T.~Reynolds$^{24}$,
D.~Ribeiro$^{1}$,
E.~Roache$^{3}$,
J.~L.~Ryan$^{22}$,
I.~Sadeh$^{15}$,
M.~Santander$^{19}$,
G.~H.~Sembroski$^{25}$,
R.~Shang$^{22}$,
D.~Tak$^{15}$,
V.~V.~Vassiliev$^{22}$,
A.~Weinstein$^{7}$,
D.~A.~Williams$^{17}$,
and 
T.~J.~Williamson$^{10}$\\
\noindent
$^1${Physics Department, Columbia University, New York, NY 10027, USA}
$^{2}${Department of Physics and Astronomy, DePauw University, Greencastle, IN 46135-0037, USA}
$^3${Center for Astrophysics $|$ Harvard \& Smithsonian, Cambridge, MA 02138, USA}
$^4${Department of Physics, Washington University, St. Louis, MO 63130, USA}
$^5${Department of Physics and Astronomy, Barnard College, Columbia University, NY 10027, USA}
$^6${Physics Department, California Polytechnic State University, San Luis Obispo, CA 94307, USA} 
$^7${Department of Physics and Astronomy, Iowa State University, Ames, IA 50011, USA}
$^8${Department of Astronomy and Astrophysics, 525 Davey Lab, Pennsylvania State University, University Park, PA 16802, USA}
$^9${School of Physics, University College Dublin, Belfield, Dublin 4, Ireland}
$^10${Department of Physics and Astronomy and the Bartol Research Institute, University of Delaware, Newark, DE 19716, USA}
$^{11}${School of Physics and Astronomy, University of Minnesota, Minneapolis, MN 55455, USA}
$^{12}${Department of Physics, California State University - East Bay, Hayward, CA 94542, USA}
$^{13}${School of Physics and Center for Relativistic Astrophysics, Georgia Institute of Technology, 837 State Street NW, Atlanta, GA 30332-0430}
$^{14}${School of Physics, National University of Ireland Galway, University Road, Galway, Ireland}
$^{15}${DESY, Platanenallee 6, 15738 Zeuthen, Germany}
$^{16}${Physics Department, McGill University, Montreal, QC H3A 2T8, Canada}
$^{17}${Santa Cruz Institute for Particle Physics and Department of Physics, University of California, Santa Cruz, CA 95064, USA}
$^{18}${Department of Physics and Astronomy, University of Utah, Salt Lake City, UT 84112, USA}
$^{19}${Department of Physics and Astronomy, University of Alabama, Tuscaloosa, AL 35487, USA}
$^{20}${Department of Physics and Astronomy, University of Iowa, Van Allen Hall, Iowa City, IA 52242, USA}
$^{21}${Institute of Particle and Cosmos Physics, Universidad Complutense de Madrid, 28040 Madrid, Spain}
$^{22}${Department of Physics and Astronomy, University of California, Los Angeles, CA 90095, USA}
$^{23}${Institute of Physics and Astronomy, University of Potsdam, 14476 Potsdam-Golm, Germany}
$^{24}${Department of Physical Sciences, Munster Technological University, Bishopstown, Cork, T12 P928, Ireland}
$^{25}${Department of Physics and Astronomy, Purdue University, West Lafayette, IN 47907, USA}

\end{document}